\documentclass[aps,prl,twocolumn,showpacs,groupedaddress]{revtex4}

\usepackage{graphicx}

\begin{document}

\title{ Doping of a One-Dimensional Mott Insulator: Photoemision and Optical 
Studies of Sr$_2$CuO$_{3+\delta}$}   
\author{T. Valla}
 \email{valla@bnl.gov}
\author{T. E. Kidd}
\author{P. D. Johnson}
\author{K. W. Kim}
 \altaffiliation[Permanent address: ]{School of Physics and Research Center for Oxide electronics, Seoul National University, Seoul 151-747, Korea}
\author{C. C. Homes}
\author{G. D. Gu}
\affiliation{ Department of Physics, Brookhaven National Laboratory, Upton, NY, 
11973-5000}
\date{\today}

\begin{abstract}
The spectral properties of a one-dimensional (1D) single-chain Mott insulator 
Sr$_2$CuO$_{3}$ have been studied in angle-resolved photoemission and 
optical spectroscopy, at half filling and with small concentrations of extra 
charge doped into the chains via high oxygen pressure growth. The single-
particle gap is reduced with oxygen doping, but the metallic state is not 
reached. The bandwidth of the charge-transfer band increases with doping, while 
the state becomes narrower, allowing unambiguous observation of separated 
spinon and holon branches in the doped system. The optical gap is not changed upon 
doping, indicating that a shift of chemical potential rather than decrease of
corelation gap is responsible for the apparent reduction of the photoemission gap.
\end{abstract}


\maketitle

1D Mott insulators have been intensively studied both theoretically and 
experimentally because of the fundamentally different physics expected in 1D 
\cite{mott}. Systems with nearly non-interacting Cu-O chains, for example, 
represent an ideal realization of a Heisenberg spin 1/2 chain \cite{Jchain}. Early 
on, it was realized that in 1D systems the elementary electronic excitations are 
not quasiparticles (QP) as in 3D systems, but collective excitations carrying 
either spin but no charge ("spinons") or the charge of an electron but no spin 
("holon") \cite{spin-hol}. If created in such a system, a physical electron (or 
hole) decays into a pair of independent excitations. A photoemission spectrum 
that measures the spectral function of a physical hole is therefore a broad 
multi-particle continuum of spinons and holons. Similarly, in inelastic neutron scattering (INS) that measures the spin-spin correlation function, a two-spinon continuum is expected instead of well-defined magnons. The boundaries of the continuum, 
representing events where one particle is left at 
rest, may still be well pronounced and it should be possible to trace the spinon and holon branches in photoemission, 
and lower and upper boundaries in INS \cite{branches,Fabian1}. While the boundaries of the continuum have been 
identified in neutron scattering \cite{neutron,Zaliznyak}, the spinon and holon 
branches have not been resolved as clearly in photoemission \cite{nopeak}.
Theoretical studies suggest that the direct 
observation of spinon-holon branches is more likely in insulating 1D materials 
\cite{Fabian-T}. These systems, with partially filled valence bands, are 
insulators due to the strong electronic correlations that forbid double 
occupancy of sites. In the "spin-charge separated" picture, charge excitations 
(holons) are gapped, while the spinons may be gapless and form a "Fermi 
surface". Gapping of the charge excitations partially suppresses blurring of the 
branches, raising the chances for their direct observation. 

In this letter, we show that the material with a single chain per unit cell, 
Sr$_2$CuO$_{3}$, may be doped with holes when grown in a high oxygen pressure. 
The single particle gap is significantly reduced with doping, while the optical 
gap remains nearly constant, the system remaining in an insulating state. 
Sharpening of the charge transfer state and its increased dispersion in doped 
samples enables unambiguous detection of the spinon and holon branches in the raw data for the 
first time.

The photoemission experiments were carried out on a high-
resolution photoemission facility based on undulator beam-line U13UB at the 
National Synchrotron Light Source with a Scienta SES-200 electron spectrometer. 
The combined instrumental energy resolution was set to $\sim 25$ meV. The 
angular resolution was better than $\pm 0.1^\circ$ translating into a momentum 
resolution of $\pm 0.0025 \AA^{-1}$ at the 15.2 eV photon energy used in the 
study. The large single crystals were grown by traveling solvent-floating zone 
method under different oxygen pressure. For photoemission studies, 
samples were mounted on a liquid He cryostat and cleaved \textit{in-situ} in the 
UHV chamber with base pressure $3\times 10^{-9}$ Pa. The polarized reflectance of freshly-cleaved crystals of 
Sr$_2$CuO$_{3+\delta}$ were measured over a wide frequency range ($\sim$30 to 
over 23,000 cm$^{-1}$, or 3 meV to $\sim$3 eV) on a Bruker IFS 66v/S Fourier 
transform spectrometer using an overfilling technique \cite{Chris}. The 
principal optical axes were determined from the anisotropic behavior of the 
phonons; a prominent copper-oxygen stretching mode at $\sim$ 540 
cm$^{-1}$ is observed only along the chain direction (\textit{b} axis). The 
optical properties were then calculated from a Kramers-Kronig analysis of the 
reflectance. 

\begin{figure*}
\includegraphics{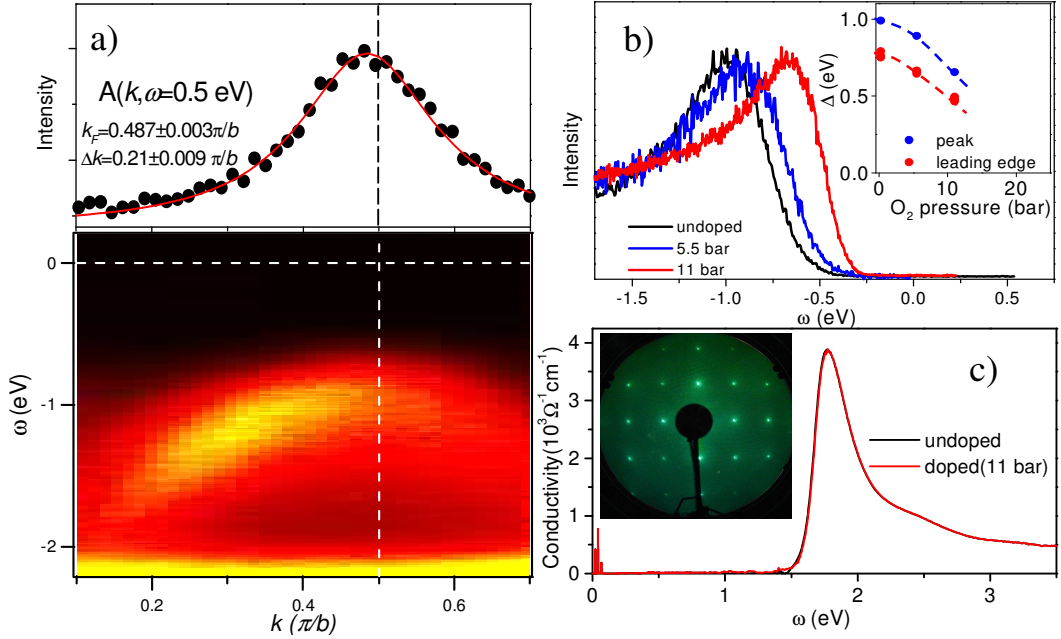}
\caption{\label{fig:1}
a) Photoemission intensity of 5.5 bar sample in the chain direction. Lower panel 
shows false color contour map with highest intensity in yellow. The dashed lines 
represent $\omega=0$ and $k=\pi/(2b)$. Upper panel represents the cross-section 
of the same spectrum (MDC) at $\omega=0.5$ eV. The red line is Lorentzian fit to the data.
b) Photoemission spectra at $k_F$ for three different growth conditions of Sr$_2$CuO$_{3+\delta}$. Spectra 
were recorded at 200K (doped) and 300 K (undoped) samples. Inset: the single-particle gap dependence on O$_2$ growth pressure.
c) Optical conductivity at room temperature for light polarized along the $b$ 
axis (chain direction) for the undoped and 11 bar O$_2$ grown samples. Inset: LEED pattern for 11 bar sample at 164 eV 
electron energy.}
\end{figure*}

Fig.~\ref{fig:1}a) shows the photoemission intensity for Sr$_2$CuO$_{3+\delta}$ grown under 5.5 bar of O$_2$, measured along the chain direction, $k_b$, with the in-plane momentum perpendicular to the chains set to $k_a=0$. 
The dispersion of the lower Hubbard band (or more correctly - the charge 
transfer band) gapped by a large correlation gap is clearly visible. The state disperses upwards for $k<k_F$, 
achieves a maximum at $k=k_F$ and then turns down for $k>k_F$. 
In Fig.~\ref{fig:1}b), the intensity at fixed $k=k_F\approx\pi/(2b)$ (the so called energy distribution curve (EDC)) is plotted for three different growth conditions. It is obvious that the gap magnitude decreases 
with the increasing O$_2$ growth pressure, from $\approx$0.75eV for the undoped 
material to $\approx$0.5eV for the highest growth pressure, measured as the position of 
the "leading edge". In the inset, we show the growth pressure dependence of the 
single-particle gap, which suggests that the metallic state might be reached for 
the O$_2$ pressure of 20-30 bar. The obvious reduction of the 
photoemission single-particle gap with increasing oxygen content is in sharp 
contrast with the behavior observed in optical measurements. 

The optical conductivity (or current-current correlation function) \cite{Kubo} represents an integral throughout the Brillouin 
zone of all the $q=0$ transitions. As can be seen in Fig.~\ref{fig:1}c), the optical gap 
remains constant upon doping. Moreover, the real part of the 
optical conductivity $\sigma_1(\omega)$ does not show any significant difference 
for samples grown under different conditions. The 
photoemission gap is approximately 1/2 of the optical gap for the undoped sample but not for the 11 bar 
grown sample. This would suggest that the 
correlation gap has not changed, but rather, the chemical potential has 
been shifted closer to the lower Hubbard band. However, the shift is pronounced only for the highest occupied state and only near $k_F$. At the zone center the state does not shift with doping. Also, the deeper states shift only by a fraction of the gap reduction.
The fact that the insulating character 
of a 1D Mott insulator is preserved upon doping is fundamentally new 
and in contrast with 2D cuprates where a small concentration (1-2\%) of doped carriers 
induces states at the Fermi level and the development of metallic in-plane transport \cite{2D}. 

\begin{figure*}
\includegraphics{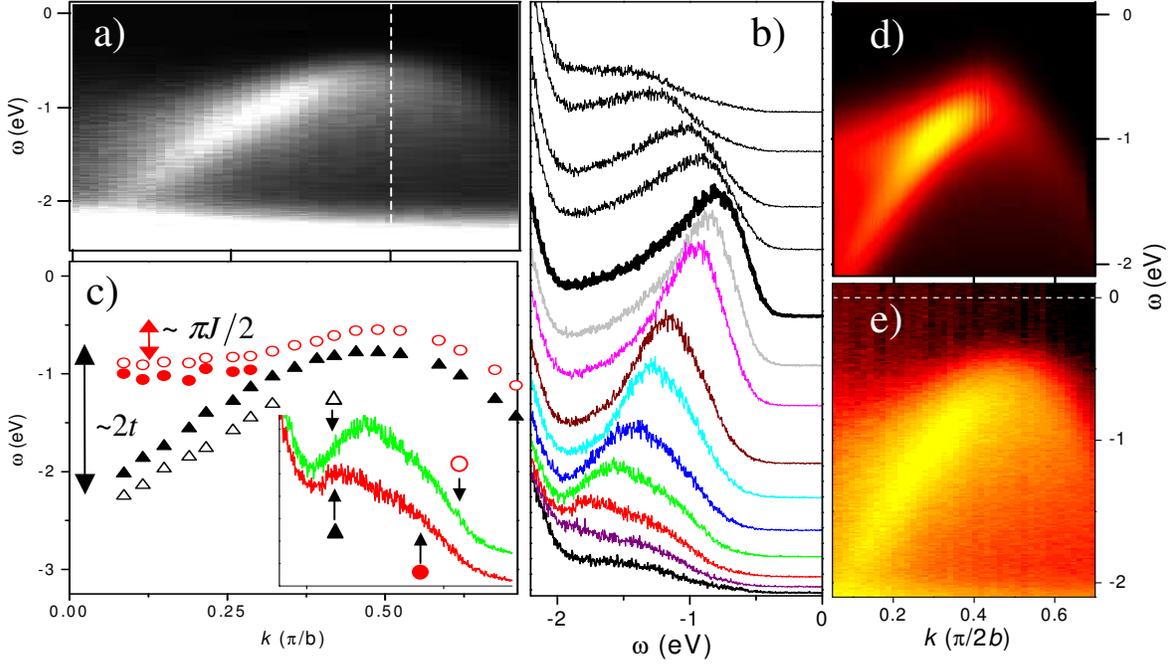}
\caption{\label{fig:2}
a) Contour plot of photoemission intensity recorded at 180 K from a freshly 
cleaved 11 bar grown sample in the chain direction. The dashed line represents $k=\pi/(2b)$. 
b) A set of corresponding EDCs for several momenta, increasing from bottom to the 
top. The thick spectrum corresponds to $k_F$. 
c) Dispersion of characteristic features in the spectra from panel (b). Solid 
(empty) circles represent "full" ("middle") step on the upper (spinon) edge, 
while triangles represent corresponding features on the lower (holon) edge, as 
indicated in the inset for re-plotted 3$^{rd}$ and 4$^{th}$ EDC from the bottom 
in panel (b).
d) The spectral function simulated as explained in the text \cite{Fabian1}.
e) The same spectrum as in a), but with intensity on a log-scale, emphasizing the spinon branch.}
\end{figure*}

Although the exact oxygen content is not known for our samples, we 
anticipate that the doping level is very low. As the structure of the
cleaved surfaces, determined by LEED (Fig.~\ref{fig:1}c), inset), is the same for all the 
samples, the system retains its orthorhombic structure for the 
range of growth conditions applied here. This is consistent with the previous 
study on lightly-doped polycrystalline samples \cite{Goodenough} where the insulating phase and the structure was preserved for $\delta<0.03$ to 0.1. That study also suggested
that the extra oxygen went into the same interstitial sites in the rocksalt 
Sr(La)O structure as in oxygen-doped La$_2$CuO$_{4+\delta}$. Such interstitial 
oxygen can oxidize the Cu-O chains in the same manner as it oxidizes the 
planes in La$_2$CuO$_{4+\delta}$. The effective doping level 
of Cu-O chains should then be directly measurable in photoemission, by measuring $k_F$. For the 
undoped material, the chains are half-filled, and $k_F$ is exactly $0.5\pi/b$. 
If the chains were doped with holes, the $k_F$ should be somewhat smaller. The upper panel in Fig. 1a) represents the 
momentum distribution curve (MDC) at energy $\omega$ set inside the gap 
where only the "tail" of the lower Hubbard band extends. The 
MDC is peaked very close to $0.5\pi/b$ for all our samples, indicating that the departure from 
half-filling is smaller than or equal to the experimental uncertainty of 2-3\% \cite{uncertainty}. 

Another intriguing feature is an apparent sharpening of the charge transfer state 
with doping (see Fig.~\ref{fig:1}b)). If the additional oxygen were to introduce disorder 
into the Cu-O chains, the effect would be opposite. It is more probable that the 
charge induced from the interstitial sites helps in screening the photohole 
created in the photoemission proces. Irrespective of the origin, the 
observed sharpening is found to be essential for detection of spin-charge separation in the 
high-pressure grown samples as discussed below.

We note that the extra oxygen tends to leave the surface area probed in 
photoemission relatively quickly after cleaving, shifting the probed layer 
closer to the undoped case. After a day in ultra-high vacuum, the spectra 
of an 11 bar grown sample look identical to the 
freshly cleaved 5.5 bar grown sample. If re-cleaved, the 11 bar characteristics 
are recovered reproducibly, meaning that only the surface region was affected. 
The loss of oxygen and its resulting inhomogeneous concentration might be a dominant factor
that limits the total photoemission width of the measured states.
Although the width and the gap increase in time, it 
is not clear what the contribution to the initial width was. However, the 
spectra of the high pressure grown samples were sharp enough to observe the 
spin-charge separation directly in the raw data. Fig.~\ref{fig:2}a) shows the 
photoemission intensity in the in-chain direction from an 11 bar sample at 
$T=180$ K, within 15 minutes after cleaving. In the raw data contour 
plot, there is already a hint of two separated branches in the dispersion of the highest 
occupied state for $k<k_F$. In panel b), the same spectrum is shown as a set of 
EDCs for several different momenta. Although the dispersion may seem symmetric 
relative to $k_F$, there is a significant difference in the lineshape between the 
$k<k_F$ and  $k>k_F$ regions. For $k<\pi/(3b)$, the spectral intensity has a 
complex shape in that it seems to be limited between two dispersing step-
like features: a rapidly dispersing lower edge and a slowly dispersing upper 
edge. As these edges merge, forming a wedge-like structure, only a single-peak 
can be identified near $k_F$. A single-peak structure remains above $k_F$. This 
is exactly what is expected if spin-charge separation takes place. The 
rapidly dispersing lower edge represents the holon branch while the upper edge 
represents the spinon branch. In the cases where these two edges are well 
separated, one can identify the "mid-point" and the "full-step" energies for 
each edge, as shown in the inset of Fig.~\ref{fig:2}c), and plot them vs. 
corresponding momenta. In cases where only a single peak is present, we plot the 
peak position and the upper (leading) edge position. The results are shown in 
panel c). If we identify the lower edge as the holon and the upper edge as the 
spinon branch, we get approximately 1.64 eV and 0.4 eV for the total holon and 
spinon bandwidth, respectively. In the $t-J$ model, this corresponds to $t=0.82$ 
eV and $J=0.26$ eV, somewhat larger than previous estimates for an undoped 
system \cite{Jchain,Zaliznyak,t-J}. As noted earlier, the dominant change induced by the 
increasing oxygen content is the gap reduction. At the same time, the bottom of 
the holon branch (at $k=0$) remains nearly at the same energy ($\sim$2.5 eV). 
That means that the holon bandwidth increases with doping. By 
using the gap dependence on pressure, we obtain $t\sim0.66$ eV and 0.72 eV for 
the undoped and 5.5 bar grown material, respectively. We were able to resolve 
the spinon branch in the 5.5 bar sample, but not in the undoped system. Compared 
to the 11 bar sample, $J$ is slightly reduced, but within the error 
of approximately $\pm0.05$ eV. 

To the best of our knowledge, our photoemission spectra from high oxygen 
pressure grown Sr$_2$CuO$_{3+\delta}$ provide the best direct evidence for spin-
charge separation so far. The continuum of excitations, with well-resolved 
spinon and holon branches, have been identified from the raw data without 
resorting to the technique of second derivatives \cite{Fujisawa,Mizokawa}. Spurious states, dispersing in 
the wrong direction near the zone center and the zone boundary from ref. 
\cite{Fujisawa}, have not been detected. 
To better illustrate the agreement with the theoretical models for 1D Mott 
insulators, we have simulated the spectral function in Fig. 2(d), by using eq. (23) from ref. \cite{Fabian1} with $v_c=6v_s$ and convoluted it with an 0.4 eV gaussian. Even though the agreement is remarkable, there are still some 
discrepancies between our data and the theoretical models. 
For example, the observed asymmetry in intensity between the holon 
and spinon branches cannot be explained in a simple 1-band Hubbard model.
Theories going beyond 1-band model seem to be able to account 
for an apparent weakening of the spinon branch \cite{3-band}. To emphasize the lower intensity spinon component, we have replotted the experimental spectrum with a log-scale for intensity in panel (e). 
An additional problem is the anomalously large 
broadening and the "leakage" into the gap that is clearly visible in the photoemission 
data (Fig.~\ref{fig:1}). As spinons are gapless, the "Fermi surface" that they 
form will be smeared out at finite temperature. However, as recent finite 
temperature calculations have shown, this will affect the holon branch much more 
strongly, while the spinon edge will remain relatively sharp \cite{Fabian-T}. Our photoemission and optical spectra show virtually no temperature dependence, with the "leakage" still present at low 
temperatures.

In conclusion, we have succeeded in doping the single chain Mott insulator 
Sr$_2$CuO$_3$ with holes via high-pressure oxygen growth. Upon doping, the 
single particle gap decreases while the optical gap remains constant. The $k_F$ 
remains close to $\pi/(2b)$, indicating a very low doping level. In 
doped samples, the separated spinon and holon branches have been detected. From 
the measured dispersions, we have deduced the set of $t-J$ parameters that agree 
well with some earlier estimates \cite{Zaliznyak,t-J} and seem to increase with doping.

The authors would like to acknowledge useful discussions with Alexei Tsvelik, 
Igor Zaliznyak, Young-June Kim and especially with Fabian Essler, who guided us 
through the 1D physics. The work was supported by the Department of 
Energy under contract number DE-AC02-98CH10886.


\begin{references}
\bibitem{uncertainty} Although the angular resoulution allows much higher precission, 
the uncertainty in the work function and/or in the absolute angle may produce a 2-3\% uncertainty in $k_F$.
\end{references}
\end{document}